\documentclass[11pt]{article}
\usepackage{moriond,epsfig}

\def\be{\begin{equation}}
\def\ee{\end{equation}}
\def\bea{\begin{eqnarray}}
\def\eea{\end{eqnarray}}

\begin{document}
\vspace*{4cm}
\title{Top Quark Properties at the TeVatron}

\author{ Yvonne Peters}

\address{ University of Manchester,   Oxford Road,    Manchester,  M13 9PL,  UK }


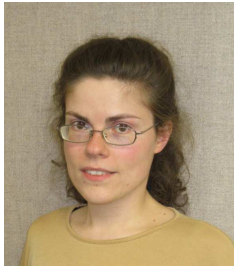
\maketitle\abstracts{Discovered in 1995 by CDF and D0 at the Fermilab Tevatron collider, the top quark remains interesting to test the Standard Model. Having collected more than 7~fb$^{-1}$ of integrated luminosity with both experiments until today, several top quark properties have been measured with increasing precision, while other properties have been investigated for the first time. 
In this article recent measurements of top quark properties from CDF and D0 are presented, using between 1~fb$^{-1}$ and $4.8$~fb$^{-1}$ of data. In particular, the measurement of the top quark mass, the top quark width, the top antitop mass difference, a check of the electric charge of the top quark,  measurements of the top antitop  quark spin correlation and W helicity as well as a search for charged Higgs bosons are discussed.
}

\section{Introduction}
 Until today, the top quark is the heaviest known elementary particle, with a top quark mass of $173.1 \pm0.6 {\rm(stat)} \pm 1.1 {\rm(syst)~GeV}$~\cite{topmassworldaverage}. This property, together with the fact that due to its short lifetime it decays before it can hadronize, makes the top quark a unique particle to study a bare quark and explore new physics. 

In order to use the top quark to search for new physics, it is crucial to understand if it is really the particle we expect it to be. In particular,  if  for example the charge, the decay properties, the lifetime or the spin correlation deviate from the values predicted by the Standard Model of particle physics~(SM), it could point to physics beyond the SM. 

Having collected about 7~fb$^{-1}$ of integrated luminosity for each experiment, the CDF and D0 detectors at the Fermilab Tevatron collider provide a large amount of data, enabling to measure many top quark properties with high precision and others for the first time. For example, the top quark mass measurement exceeds 1\% of accuracy now in a single measurement. 
In the following, recent measurements of several top quark properties measured by CDF and D0, including a search for charged Higgs bosons in top quark decays, are presented. 

\section{Top Quark Mass, Width and Top Antitop Quark Mass Difference}
The top quark mass is not predicted by the SM. Besides the interest in itself to determine this free parameter of the SM, the top quark mass, together with the $W$ boson mass, constrains the mass of the not yet found  Higgs boson. 

\subsection{Top Quark Mass}
Several techniques have been developed to determine the top quark mass as precise as possible. The simplest method consists of the construction of templates that depend on the top quark mass, which can be fitted to the data. A recent measurement using this technique in the lepton plus jets and dilepton final states~\cite{productionalison} reconstructs the top quark mass from the two $b$-quark jets, the light jets and the leptons. The neutrino energy is reconstructed using constraints on the mass of the $W$ boson in lepton plus jets events. 
In case of dileptonic events, the kinematics is underconstraint due to the two neutrinos in the final state which leave the detector without interacting, and an additional integration over unknown quantities is performed. 
Using $4.8$~fb$^{-1}$ of data, the most recent result from CDF is $m_t=171.9 \pm 1.1{\rm (stat+JES)}\pm0.9{ \rm (syst)~GeV}$~\cite{masstemplatecdf}. The uncertainty due to Jet Energy Scale (JES) is the dominating systematic uncertainty, but can be constraint from the hadronically decaying $W$ boson in lepton plus jets events, resulting in a reduced dependency of the top quark mass measurement on the JES uncertainty.

Another important technique to measure the top quark mass is the so-called Matrix Element (ME) technique. This method is the most precise one, as it uses the full kinematic information of the event. The basic idea is the calculation of a per-event probability $P_{sig}(x;m_t)$, where $x$ denotes the momenta of the final state partons. $P_{sig}$ is the integral over the parton distribution functions, folded with a leading order (LO) Matrix Element for the $t\bar{t}$ production and transfer functions that describe the transition of the true momenta into the measured momenta of the final state particles coming from the top quark decays. Finally, the likelihood of the product of these event probabilities is minimized, yielding the measured cross section. By comparing input and measured top quark masses in ensemble tests, the method is validated and has to be calibrated for a small bias, due to only using LO matrix elements. More details can be found in Refs.~\cite{massljetsmed0,massljetsmecdf}.  The most recent measurements using this technique in the lepton plus jets final state result in $m_t=173.7 \pm 1.3{\rm (stat+JES)}\pm1.4{ \rm (syst)~GeV}$ with $3.6$~fb$^{-1}$ of D0 data~\cite{massljetsmed0} and  $m_t=172.8 \pm 0.9{\rm (stat+JES)}\pm0.8{ \rm (syst)~GeV}$ with $4.8$~fb$^{-1}$ of CDF data~\cite{massljetsmecdf}, the latter being the most precise single measurement of the top quark mass today.

Since the discovery of the top quark, several top quark mass measurements in various channels and using various techniques have been performed at the Tevatron. Combining all of these measurements a precision of $0.75\%$ has been achieved on the top quark mass~\cite{topmassworldaverage}, not yet including the two recent measurements form CDF shown here.

Besides the direct measurement, recently the top quark mass has been extracted from the measurement of the $t\bar{t}$ cross section~\cite{xseccombiandmass}, resulting in $m_t=169.1^{+5.9}_{–5.2} {\rm (stat+syst)~GeV}$. Although this result is much less precise than the direct mass measurements, it can clearly be interpreted as a measurement of the pole mass of the top quark, since inclusive cross section calculations are used that are performed in the pole mass scheme. In contrast to this it is not exactly clear which effective mass scheme is used in Monte Carlo simulations used for the direct measurements of the top quark mass~\cite{massproblem,masstheorymoriond}.

\subsection{Top Antitop Quark Mass Difference}
In the SM, particles and their corresponding anti-particles have identical masses. If this were not the case, it would indicate CPT violation. 
By extending the event probabilities $P_{sig}(x;m_t)$ to $P_{sig}(x;m_t, m_{\bar{t}})$, the Matrix Element method can be extended from a top quark  mass measurement to a measurement of the top antitop quark mass difference. Using $1$~fb$^{-1}$ of data, the D0 collaboration performed the first measurement of a mass difference between a  bare quark and its antiquark. The analysis was performed in the lepton plus jets channel, resulting in  $m_t - m_{\bar{t}} = 3.8\pm3.7$~GeV~\cite{massdifference}, consistent with the SM. This measurement is still dominated by the statistical uncertainty, making it interesting to be repeated with more data.

\subsection{Top Quark Width}
The top quark lifetime is known to be short in the SM, resulting in a top quark width of about $\Gamma_t = 1.4$~GeV at the world average top quark mass. Using the lepton plus jets final state, CDF has extended its mass measurement using the template method to extract information on the top quark width. The templates are the same as for the top quark mass measurement, i. e. the reconstructed top quark mass, but instead of varying the mass, the top quark width is fitted to the data. Using $4.3$~fb$^{-1}$ of data, CDF determines the top quark width to be $0.4 < \Gamma_t<4.4$~GeV at 68\% confidence level (CL), or $\Gamma_t<7.5$~GeV at 95\% CL~\cite{topwidth}, in agreement with the SM prediction.

\section{Top Quark Charge}
In the SM, the charge of the top quark is predicted to be that of a normal up-type quark, i. e. $+2/3$ of the electron charge. Nevertheless, it is possible that the top quark has an exotic charge of $-4/3$e. Distinguishing the two requires the measurement of the top quark charge in data. 
A recent charge measurement from CDF using events in the lepton plus jets final state exploits soft lepton $b$-tagging to test the SM and the exotic charge hypotheses. The event selection requires at least one jet to be $b$-tagged with a secondary vertex algorithm, and at least one jet (which is allowed to be the same one) to contain a soft lepton, coming from a semileptonic $B$ decay. Furthermore, a kinematic fit is performed, where final state particles from the top and antitop quark decay are assigned together using constraints from the $W$ boson mass, the requirement of the top quark mass and finally a low $\chi^2$ of the fit. 
Having then two leptons -- one from the $W$ boson decay and one from the $b$-jet -- with clear charge information, the knowledge from the kinematic fit allows to predict if both leptons come form the same top quark or one from the top and one from the antitop, and thus yield information about the top quark charge. With this technique the exotic top quark charge of $-4/3$e can be excluded with 95\% CL~\cite{chargecdf}. Somewhat older top quark charge measurements from CDF~\cite{chargeoldcdf} and D0~\cite{charged0}, performed with a slightly different technique, also prefer the SM top quark charge, with 87\% and 92\% CL respectively.

\section{W Helicity Fractions}
The coupling of $W$ bosons to fermions is purely left-handed in the SM, thus the relative orientation of the spin of the $b$-quark and the $W$ boson from the top quark decay are known. The fraction of negative ($f_{-}$), zero ($f_0$) and positive ($f_{+}$) helicity of the $W$ boson are predicted to be about 30\%, 70\% and less than $3.6\times10^{â-4}$, respectively. In order to measure the $W$ boson helicity fractions, the angle $\theta^{*}$ between the down-type decay product of the $W$ boson in the rest frame of the latter, with respect to the $W$ boson flight direction in the top quark rest frame is measured. The distribution of the cosine of this angle differs for the three possible helicity fractions, which is exploited in the measurement.

Similar to the top quark mass, several techniques are used to determine the $W$ boson helicity fractions. The simplest one is again a template method, where the $\cos(\theta^{*})$ distribution is fitted to the data. To be as model independent as possible, the fractions $f_0$ and $f_{+}$ are fitted simultaneously. Using this method and $2.7$~fb$^{-1}$ of data, D0 measured $f_0=0.49\pm.011{\rm(stat)}\pm0.09{\rm (syst)}$ and $f_{+}=0.11\pm0.06{\rm(stat)}\pm0.05{\rm(syst)}$~\cite{whelicityd0}, in agreement with the SM.

The other method is the Matrix Element method, where again event probabilities are calculated. In this case $P_{sig}$ depends on $f_0$ and $f_{+}$ instead of the top quark mass, with the top quark mass fixed to $175$~GeV. A recent result form CDF applying the ME technique using $2.7$~fb$^{-1}$ of data yields $f_0=0.88\pm.011{\rm(stat)}\pm0.06{\rm (syst)}$ and $f_{+}=-0.15\pm0.07{\rm(stat)}\pm0.06{\rm(syst)}$~\cite{whelicitycdf}, also in agreement with the SM.

\section{\bf \boldmath $t\bar{t}$ Spin Correlations}
One of the special features of the top quark besides its high mass is its short lifetime,  shorter than the timescale for hadronization. Therefore, top quarks decay before they hadronize. This implies that the spin information of the top quark is carried into its decay products, enabling the measurement of the spin correlation of the top and antitop quark in $t\bar{t}$ events.

The doubly differential cross section $1/\sigma d^2\sigma/(d\cos(\theta_1)d\cos(\theta_2))$ can be written as $1/4(1-C\cos(\theta_1)\cos(\theta_2))$, where $C$ is the spin correlation strength, and $\theta_1$ ($\theta_2$) is the angle of the down-type fermion from the $W$ boson or the top (antitop) quark decay in the top (antitop) rest frame, with respect to a quantization axis. Two common choices for the quantization axis are the top (antitop) flight direction in the $t\bar{t}$ rest frame -- the so-called helicity basis -- or the beam axis, in the so-called beam basis. Depending on the quantization axis and the collision energy, the SM values of $C$ are different.
The spin correlation strength $C$ can be considered as the number of events with top and antitop having the same spin direction minus the number of events with opposite spin direction, normalized to the total number of $t\bar{t}$ events, multiplied with a factor representing the analyzing power of the down-type fermion used to calculate the angles. For leptons this additional factor is one, while it is smaller for quarks. This latter fact results in the spin correlation analysis to be more sensitive in dilepton events although the statistics is lower than in  lepton plus jets events. Moreover, the information in dilepon events is lower, as the two neutrinos are not identified in the detector.

The outline of the measurement of spin correlations in $t\bar{t}$ events is the following: The distribution of $\cos(\theta_1) \cos(\theta_2)$ looks different in case of $C=0$ relative to the SM value of $C$. In dilepton events, CDF and D0 performed an analysis of the spin correlation strength using $2.8$~fb$^{-1}$ and $4.2$~fb$^{-1}$ of data, respectively. The CDF measurement using the off-diagonal basis yields $C=0.32_{-0.78}^{+0.55} {\rm(stat+syst)}$~\cite{spincorrelationllcdf}, and the D0 analysis yields $C=-0.17_{-0.53}^{+0.65}{\rm(stat+syst)}$ in the beam basis~\cite{spincorrelationlld0}, in both cases in agreement with a next to leading order (NLO) SM value of $C \simeq 0.78$~\cite{Bernreuther:2004wz}. 

Recently, CDF performed the first extraction of the $t\bar{t}$ helicity fractions in lepton plus jets events. Similar to the analyses in dileptonic final states, templates of equal and opposite $t\bar{t}$ helicity are used and fitted to the data. The measured quantity can then directly be translated into the spin correlation strength $C$ in the helicity basis. With a dataset of $4.3$~fb$^{-1}$, the spin correlation in lepton plus jets events is $C=0.60\pm0.50{\rm(stat)}\pm0.16{\rm(syst)}$~\cite{spincorrelationljetscdf}, consistent with a NLO SM value of $C=0.35$ in the helicity basis~\cite{Bernreuther:2004wz}.

All these three measurements are still statistically limited.

\section{Search for Charged Higgs Bosons in Top Quark Decays}
In the SM, the top quark decays almost always into a $W$ boson and a $b$-quark. In several models, for example supersymmetric models, the Higgs sector has to be extended by at least one additional Higgs douplet,  resulting in additional physical Higgs bosons, one of which is a charged Higgs boson. In case of this charged Higgs boson being lighter than the top quark, the   decays $t\rightarrow W^{+}b$ and $t\rightarrow H^{+}b$ can compete. Over the last years, CDF and D0 have performed a variety of charged Higgs boson searches in top quark decays, where the charged Higgs boson could decay either into a tau and the corresponding neutrino, a charm and a strange (anti)quark or into a $W$ boson and a CP-odd neutral Higgs boson $\mathcal A$. 

Several search strategies are used. There are searches for a signal in a specific final state. Furthermore, the existence of a charged Higgs boson can change the fractions between different final states. Searches using both strategies are presented here. 

Assuming $B(H^{+}\rightarrow c\bar{s})=100\%$, CDF performed a search in the lepton plus jets channel using the invariant mass of the non-$b$-jets. For charged Higgs boson masses lighter or heavier than the $W$ boson mass, a peak in the invariant mass spectrum would give hints for a signal. As no sign of new physics could be seen, limits on $B(t\rightarrow H^{+}b)$ for $m_{H^{+}}$ between 60 and 70~GeV and between 90 and 150~GeV were set, varying between $B<0.08$ and $B<0.32$ at 95\% CL~\cite{chargecdf} for a dataset of $2.2$~fb$^{-1}$.

Another recent search for charged Higgs bosons from CDF in the lepton plus jets channel assumes $B(H^{+}\rightarrow W^{+}{\mathcal A})=100\%$, where the mass of the CP-odd neutral Higgs boson $\mathcal A$ is below two times the $b$-quark mass. In the next to minimal supersymmetric model this scenario is still compatible with the limits on neutral Higgs bosons form LEP~\cite{lephiggs}.
It is assumed that  $\mathcal A$ decays exclusively into two tau leptons. The distribution of low $p_{T}$ isolated tracks from the tau lepton is investigated for a possible signal from charged Higgs bosons. As no hints for a signal can be seen, limits on $B(t\rightarrow H^{+}b)$ for $m_{H^{+}}$ between 80 and 160~GeV are set for various masses of the CP-odd neutral Higgs boson. These limits, ranging from 65\% down below 10\% (depending on the charged Higgs and neutral Higgs boson mass)  for $2.7$~fb$^{-1}$ of data, are the first limits for such a scenario from the Tevatron~\cite{hplusnmssmcdf}.

A further search strategy explores the different distribution of events between various final states compared to the SM, in case the decays $t\rightarrow W^{+}b$ and $t\rightarrow H^{+}b$ occur both. By comparing events in the lepton plus jets, dilepton and $\tau+$lepton final states, D0 performed a search for charged Higgs bosons using the constraint $B(H^{+}\rightarrow \tau \nu)+B(H^{+}\rightarrow c \bar{s})=100\%$. Assuming the charged Higgs boson decays tauonically, the number of events is expected to decrease in the lepton plus jets and dilepton final states, but to increase in the $\tau+$lepton channel for increasing $B(t\rightarrow H^{+}b)$. A fit to the number of events in the different decay channels is performed, yielding limits on $B(t\rightarrow H^{+}b)$. To decrease the systematic uncertainties a simultaneous fit to both the branching ratio $B(t\rightarrow H^{+}b)$ and the $t\bar{t}$ cross section is performed. Using this approach, upper limits on $B(t\rightarrow H^{+}b)$ between 0.13 for low charged Higgs masses and
0.26 for high masses at 95\% CL can be set for pure tauonic decays of the  charged Higgs bosons. For hadronically decaying charged Higgs bosons, upper limits of 0.22 at 95\% CL can be set, using  $1$~fb$^{-1}$ of data~\cite{Hplus_global}. 

These limits on $B(t\rightarrow H^{+}b)$ can be used to exclude regions in the parameter space $[M_{H^{+}},\tan \beta]$ of the minimal supersymmetric standard model, where $\tan \beta$ is the ratio of vacuum expectation values of the two Higgs duplets. For the search from D0 different benchmark scenarios were considered, e. g. a CP-violating model. For the latter a strangephilic CPX scenario~\cite{strangephilicCPX} was used, where the charged Higgs boson decay into charm and strange quarks is enhanced by introducing a hierarchy between the first two and the third generation of sfermions. Charged Higgs boson masses up to 154~GeV can be excluded for large $\tan \beta$ in this model.

\section{Conclusion and Outlook}
In this article, a collection of recent measurements of top quark properties has been presented. Up to $4.8$~fb$^{-1}$ of data collected by the CDF and D0 detector was used, which allows precision measurements of top quark properties. Until today, more than 7~fb$^{-1}$ of data have been collected, and more than 10~fb$^{-1}$ of data are expected at the end of Run~II of the Tevatron. Combined with the ever improving techniques, the better understanding of our data and the increasing statistics, the top quark is and will stay an interesting particle to study.

\section*{Acknowledgments}
I would like to thank my collaborators from the CDF and D0 collaborations for their help in
preparing this article. I also thank the staffs at Fermilab and collaborating institutions, and acknowledge the support from the Royal Society.


\section*{References}
\begin{thebibliography}{99}
\bibitem{topmassworldaverage}    Tevatron Electroweak Working Group and CDF Collaboration and D0 Collab,
  arXiv:0903.2503 [hep-ex].

\bibitem{productionalison} Details about $t\bar{t}$ production and final states: A. Lister, these proceedings.


\bibitem{masstemplatecdf} The CDF Collaboration, ``Combined template-based Top Quark Mass Measurement in the Lepton+Jets and 
Dileptons Channels Using $4.8$~fb$^{-1}$ of data'', CDF note 10033. 


\bibitem{massljetsmed0} The D0 collaboration, ``Measurement of the Top Quark Mass in the Lepton+Jets Channel Using the Matrix 
Element Method on $3.6$~fb$^{-1}$ of D0 Run II Data'', D0 note 5877-CONF. 

\bibitem{massljetsmecdf} The CDF Collaboration, ``Top Mass Measurement in the Lepton + Jets Channel Using a Matrix 
Element Method with Quasi-Monte Carlo Integration and in situ Jet 
Calibration with $4.8$~fb$^{-1}$, CDF note 10077.

\bibitem{xseccombiandmass} V.~M.~Abazov {\it et al.}  [D0 Collaboration], 
 Phys.\ Rev.\  D {\bf 80}, 071102 (2009).

\bibitem{massproblem}
  A.~H.~Hoang and I.~W.~Stewart,
  Nucl.\ Phys.\ Proc.\ Suppl.\  {\bf 185}, 220 (2008).

\bibitem{masstheorymoriond} Peter Uwer, these proceedings.

\bibitem{massdifference}  V.~M.~Abazov {\it et al.}  [D0 Collaboration],
 Phys.\ Rev.\ Lett.\  {\bf 103}, 132001 (2009).


\bibitem{topwidth} The CDF Collaboration, 
``A Measurement of Top Quark Width using Template Method in Lepton+Jets 
Channel with $4.3$~fb$^{-1}$'', CDF note 10035. 

\bibitem{chargecdf} The CDF Collaboration, ``Exclusion of Exotic Top-like Quark with 
−4/3 Electric Charge Using Soft Lepton Tags'', CDF note 9939. 

\bibitem{chargeoldcdf} The CDF Collaboration, ``First CDF Measurement of the Top Quark Charge using the Top Decay Products'', CDF note 8967. 

\bibitem{charged0}
 V.~M.~Abazov {\it et al.}  [D0 Collaboration],
  Phys.\ Rev.\ Lett.\  {\bf 98}, 041801 (2007).

\bibitem{whelicityd0} The D0 Collaboration, ``Model-independent measurement of the W boson helicity in top quark decays at D0'', D0 note 5722-CONF.


\bibitem{whelicitycdf} The CDF Collaboration, ``Measurement of W Boson Helicity Fractions in Top Quark Decay to Lepton+Jets Events using a Matrix Element Analysis Technique with $2.7$~fb$^{-1}$ of Data'', CDF note 10004.



\bibitem{spincorrelationllcdf} The CDF Collaboration, ``A Measurement of $t\bar{t}$ Spin Correlations Coefficient in $2.8$~fb$^{-1}$ Dilepton Candidates'', CDF Conf-note 9824. 

\bibitem{spincorrelationlld0} The D0 Collaboration, ``Spin correlations in $t\bar{t}$ production in dilepton final states'', D0 note 5950-CONF. 

\bibitem{Bernreuther:2004wz}
  W.~Bernreuther, A.~Brandenburg, Z.~G.~Si and P.~Uwer,
  arXiv:hep-ph/0410197.


\bibitem{spincorrelationljetscdf} The CDF Collaboration,``Measurement of $t\bar{t}$ Helicity Fractions 
and Spin Correlation Using Reconstructed Lepton+Jets Events'', CDF note 10048.



\bibitem{hpluscdf}
  T.~Aaltonen {\it et al.}  [CDF Collaboration],
  Phys.\ Rev.\ Lett.\  {\bf 103}, 101803 (2009).

\bibitem{lephiggs} The LEP Higgs working group, Phys. Lett. B {\bf 565}, 61 (2003).


\bibitem{hplusnmssmcdf} The CDF Collaboration, 
``Search for Light Higgs Boson from Top Quark Decay'', CDF note 10104.

\bibitem{Hplus_global}
  V.~M.~Abazov {\it et al.}  [D0 Collaboration],
  Phys.\ Lett.\  B {\bf 682}, 278 (2009).
                                                                                                                                                                                                                                               
\bibitem{strangephilicCPX}
  J.~S.~Lee, Y.~Peters, A.~Pilaftsis and C.~Schwanenberger,
  Eur.\ Phys.\ J.\  C {\bf 66}, 261 (2010).


\end{thebibliography}
\end{document}